\documentclass[twoside]{dis04}
\def\runauthor{R.S. Thorne {\it et al.}}
\def\shorttitle{Update of MRST Parton Distributions}

\def\be{\begin{equation}}    
                                                      
\def\ee{\end{equation}}                                                         
\def\bea{\begin{eqnarray}}     
                                                     
\def\eea{\end{eqnarray}}
\def\GeV{{\rm GeV}} 

\begin{document}

\title{Update of MRST Parton Distributions}

\author{R.S. Thorne}

\address{Cavendish Laboratory, University of Cambridge\footnote{Royal 
Society University Research Fellow}, Madingley Road,\\
Cambridge, CB3 0HE, U.K.}

\author{A.D. Martin and W.J. Stirling}

\address{Department of Physics, University of Durham, Durham, DH1 3LE, U.K.}

\author{R.G. Roberts}

\address{Rutherford Appleton Laboratory, Chilton, Didcot, Oxon, OX11 0QX, 
U.K.}  

\maketitle

\abstracts{I present a brief summary of the most important updates in the MRST 
determinations of the NLO and NNLO parton distributions. I concentrate
on uncertainties due to input assumptions and possible theoretical corrections,
particularly isospin violation, QED corrections, NNLO and corrections at low 
$Q^2$ and small $x$.}

\section{Introduction} 

A knowledge of the partonic structure of the proton is an essential 
ingredient in the analysis of hard scattering
data from $pp$ or $\bar pp$ or $ep$ high energy collisions. 
Much attention has recently been devoted to obtaining 
reliable uncertainties on the parton
distributions obtained from global or semi-global fits. 
One obvious uncertainty is due to the 
systematic and statistical errors of the
data, i.e the {\em experimental} errors. They have been estimated 
by several groups \cite{Botje}--\cite{MRST2002}, working within a 
NLO (and sometimes NNLO) framework using a variety of different procedures. 
The general conclusion is that in all approaches the uncertainties are 
of the order of $2-5\%$, except for certain partons and regions of phase space,
where the uncertainty can be bigger,
particularly the high-$x$ gluons and very high-$x$ down quarks. However, when
the partons and predictions for physical quantities using different parton 
sets 
are compared, the deviations can be significantly greater than the quoted 
errors. A recent summary can be found in \cite{lp03}. This unfortunate 
discrepancy is likely to be due to different assumptions made by different
groups when performing their fitting procedures, 
such as different data sets used, parameterizations for input sets, 
choices of heavy flavour prescriptions, ordering of perturbative series,
etc. However, the different results for partons obtained when making different 
assumptions also suggest that the standard fixed order in the perturbative 
expansion is not completely adequate to describe the data, with 
different individual data sets making different and conflicting demands
on the partons and on $\alpha_S$. Indeed, MRST do see certain difficulties in 
obtaining the best possible fit to the data with the current theoretical 
treatment -- the HERA and NMC data in the range $x\sim 0.01-0.001$ are 
too steep in $Q^2$; the gluon is negative (or at best valencelike) at 
small $x$ and $Q^2$; and the Tevatron jet data prefer a different shape 
gluon from the DIS data. 
Hence, in order to understand the uncertainties in parton distributions, 
it is vital to consider theoretical corrections. These include 
the possibility of isospin violation; $s(x) \not= \bar s(x)$; 
higher orders (NNLO); QED (comparable to NNLO ? i.e. 
$\alpha_s^3 \sim \alpha$); 
large $x$ ($\alpha_s^n \ln^{2n-1}(1-x)$); low $Q^2$ 
(higher twist); small $x$ ($\alpha_s^n \ln^{n-1}(1/x)$). 
The uncertainties associated with many of these possible sources have 
recently been considered \cite{MRSTerror2}. Here I will highlight some of the
main conclusions and some additional results which postdate this analysis.

One of the issues we have examined is the possibility of isospin violation.
This is particularly relevant because of the ``NuTeV anomaly'' 
\cite{NuTevtheta}. NuTeV measure, with a $3-\sigma$ discrepancy, 
$$R^-=\frac{1}{2}-\sin^2 \theta_W +(1-\frac{7}{3}
\sin^2 \theta_W) \frac{[\delta U_{\rm v}] -[\delta D_{\rm v}]}{2[V^-]}.$$
where $[\delta U_{\rm v}] = [U^p_{\rm v}] - [D^n_{\rm v}]$ and 
$[\delta D_{\rm v}] = [D^p_{\rm v}] - [U^n_{\rm v}]$, $[U^p_{\rm v}]=
\int_0^1 x\,u^p_{\rm v}(x)\, dx$ etc..
We parameterize the isospin violation by
$$u^p_{\rm v}(x) = d^n_{\rm v}(x) + \kappa f(x), \qquad \qquad 
d^p_{\rm v}(x) = u^n_{\rm v}(x) - \kappa f(x),$$
where $f(x)$ is a number conserving function with valence like behaviour 
at large and small $x$ \cite{MRSTerror2}.
The variation of the $\chi^2$ for the global fit is shown in Fig. 1, and 
the best fit leads to $\kappa = -0.2$ (which agrees very well with a 
prediction in \cite{londergan}), and to a  
reduction of the NuTeV anomaly to $\sim 1.5-\sigma$, 
i.e. $\Delta sin^2\theta_W \sim -0.002$. However, a wide variation 
in $\kappa$ is possible,
and the anomaly may be reduced further or enlarged. 
Other issues to which the anomaly is sensitive are the 
strangeness asymmetry in the proton recently investigated in \cite{strange}
and electroweak corrections \cite{ewnutev}. In each case the uncertainties 
may be large as the anomaly, though there is some disagreement
regarding the strangeness asymmetry.

\begin{figure}[!thb]
\vspace*{4.7cm}
\begin{center}
\includegraphics{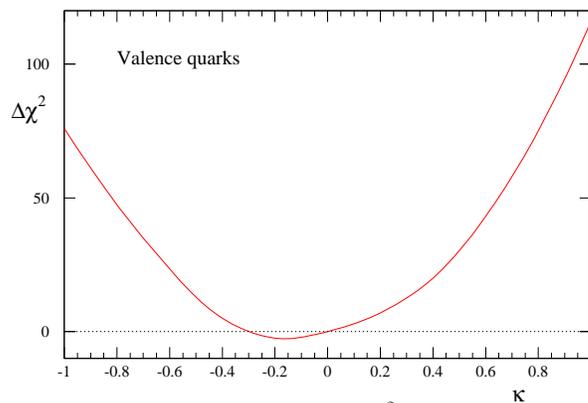}
\caption[*]{Variation of global $\chi^2$ with the value of $\kappa$}
\end{center}
\end{figure}

\vspace{-1cm}

We have also recently examined the effects due to QED.
The corrections to DIS include those in Fig. 2 and
lead to mass singularities when $\gamma \parallel q$.
The QED collinear singularities are 
{\it universal} and can be absorbed 
into pdfs, exactly as for
QCD collinear singularities, leaving finite 
(as $m_q \to 0$) ${\cal O}(\alpha)$ QED corrections
in coefficient functions, and QED--improved DGLAP equations.
The effect on the gluon distribution is negligible 
as is that on the quark distributions at small $x$ where the gluon contribution
dominates DGLAP evolution.
Only at large $x$ is the effect noticeable (${\cal O}(1\%)$) 
as first noted in \cite{spiesberger}, and more recently verified in  
\cite{Roth}, and it is equivalent to a slight
shift  in $\alpha_S$: $\Delta \alpha_S(M_Z^2)  \simeq + 0.0003 $.
This is much smaller than many sources of uncertainty.
\begin{figure}[!thb]
\vspace*{4.2cm}
\begin{center}
\includegraphics{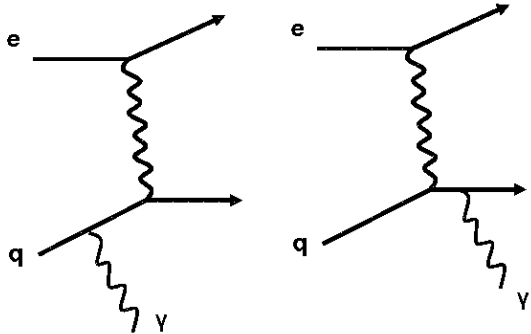}
\caption[*]{Diagrams for QED corrections to DIS}
\end{center}
\end{figure}

\vspace{-1cm}

However, these QED effects, although small, lead to isospin violation.
$u_V^p(x)$ quarks radiate more photons than $d_V^n(x)$ 
quarks. To a rough approximation, the photon distribution should be  
$$
\gamma(x,Q^2) = \sum_j e_j^2 {\alpha
\over 2\pi} \ln(Q^2/m_q^2) \int_x^1 \frac{dy}{y} 
P_{\gamma q}(y)\; \; q_j(\frac{x}{y},Q^2).
$$
So there is more photon momentum in the proton than in the neutron due to 
high-$x$ up quarks radiating more than high-$x$ down quarks. 
Momentum conservation leads to $u_V^p(x)< d_V^n(x)$ at high $x$. 
Hence, $[\delta U_{\rm v}]<0$ as required by the NuTeV anomaly and as 
found by our best fit above.
Our estimates imply that the QED effects lead to $\sim 3/4$ of the isospin 
violation observed by our best fit. Details of this and parton distributions 
involving QED corrections will soon appear \cite{MRSTQED}.  

Much more generally important are NNLO corrections. 
The coefficient functions for $F_2(x,Q^2)$ are known at NNLO \cite{CF}.
Singular limits $x \to 0$ and large $N_F$ limits are known for the NNLO
splitting functions \cite{BFKL}--\cite{SF47} 
as well as limited moments \cite{SF},  
and approximate NNLO splitting functions were  
devised \cite{VV12} and used by MRST \cite{MRSTNNLO} (and Alekhin 
\cite{Alekhin}) in NNLO fits. 
Compared to NLO they improved 
the quality of the fit very slightly,  and reduced $\alpha_S$ to $\sim 0.1155$.
The predictions lead to reasonable stability order by order for 
(quark-dominated) $W$ and $Z$ cross-sections. 
However, this fairly good convergence is largely guaranteed because 
the quarks are fit directly to data, and it is  much worse for 
gluon-dominated quantities e.g. $F_L(x,Q^2)$ or Higgs production.
Very recently, the full calculation of the NNLO splitting functions was 
completed \cite{NNLOSF}. We have confirmed that these alter the parton 
distributions obtained from the previous best estimates by only a very small 
amount. Full NNLO partons will appear soon, but also require the 
complete incorporation of NNLO coefficient functions for as many processes
as possible, and a rigorous treatment of heavy flavour thresholds at NNLO
\cite{MRSTNNLOnew}. 
For predictions it will be useful to have the exact NNLO (${\cal O}
(\alpha_S^3)$) longitudinal coefficient functions, and these, along with the 
${\cal O}(\alpha_S^3)$ (NNNLO) coefficient functions for $F_2(x,Q^2)$
will soon be complete. We have estimated the most likely value, along with 
uncertainties, from the limited information, as had previously been done for 
the splitting functions. The central value and the two ``extremes'' 
for the dominant contribution, 
i.e. the part of $C^3_{Lg}$
which is $\propto N_f$ are respectively (the common 
factor of $N_f(\alpha_S/(4\pi))^3$ 
is taken out and $L=\ln(1/x)$, $L_1=\ln(1-x)$)
\bea
& &410\frac{L}{x} -\frac{1212}{x} 
-1685 -54161x - 47131L_1 -11496L_1^2 
-867L_1^3,\nonumber\\
& &410\frac{L}{x} -\frac{0}{x} +20298 -106974x 
-67527L_1 -16336L_1^2 - 1241L_1^3 -14116L,\nonumber\\
& &410\frac{L}{x} -\frac{2057}{x} +30895x^3 +
6220L_1 -102L_1^2-66.6L_1^3+1378L^2\nonumber.
\eea 

The failure of the NNLO corrections to solve completely the problems with 
the quality of the fit leads us to try an alternative approach.
In order to investigate the real quality of the fit and the regions with 
problems we vary kinematic cuts on data, i.e. change $W^2_{cut}$, $Q^2_{cut}$ 
and $x_{cut}$, re-fit  and see if the quality of the fit to the remaining 
data improves and/or the input parameters change dramatically. 
This is continued until the quality of the fit and the partons stabilize.
This is discussed in detail in \cite{MRSTerror2}. Briefly, our $W^2$ cut
of $12.5 \GeV^2$ seems reasonable, whereas the fit improves gradually until
$Q^2_{cut}\sim 10\GeV^2$ and $x_{cut}\sim 0.005$.\footnote{We note that CTEQ
do not seem to reproduce these conclusions \cite{Tung}.} 
This is true at both NLO 
and NNLO, but the modification of the partons with changing cuts is much 
greater at NLO. For both NLO and NNLO we have defined ``conservative''
partons \cite{MRSTerror2} which are obtained using only the data within the
conservative cuts.

In order to investigate the possible theoretical improvements which could 
modify the partons at small $x$ and low $Q^2$ (particularly at high $x$)
we have tried including known $\ln(1-x)$ resummations, saturation corrections,
and purely phenomenological higher twist and $\ln(1/x)$ corrections. 
At high $x$, higher orders in the coefficient functions and higher twist 
contributions can improve the fit, but are difficult to distinguish. At low 
$x$, there is no evidence for phenomenological higher twist. Saturation 
corrections \cite{MQ} 
do not seem to help the NLO\footnote{A recent joint analysis 
with diffractive data suggests some slight improvement \cite{Watt}.} 
or NNLO fits (they would help only at LO). There
is a distinct improvement in the fit quality when (momentum conserving)
$\ln(1/x)$ corrections are added to the splitting functions, increasing the 
small-$x$ evolution. Hence, there is some reason for uncertainty
about the 
correct approach, and hence the correct predictions at very small $x$. 
Indeed, a comparison of the LO, NLO, approximate NNLO and a resummed 
prediction \cite{RST} for $F_L(x,Q^2)$ is shown in Fig. 3, exhibiting the 
level of uncertainty. Also shown are potential data which could be obtained 
by low energy running at HERA. It is clear that such data would play a 
huge role in enabling us to understand and be confident of predictions 
involving small $x$ partons. 

In summary, as well as the uncertainties on partons due to experimental
errors, there are uncertainties on partons, and related physical quantities,
due to assumptions in the fit and having to work within an incomplete 
theoretical framework. Some of the uncertainties are small, but in some 
cases, e.g. QED corrections or $s(x) \not= \bar s(x)$, they can be 
important for 
specific quantities. Others, associated with low $Q^2$ and small $x$, may
be potentially large, and more theoretical work is needed, along with  
specifically targeted data to guide and verify this work.

\begin{figure}[!thb]
\vspace*{10.2cm}
\begin{center}
\includegraphics{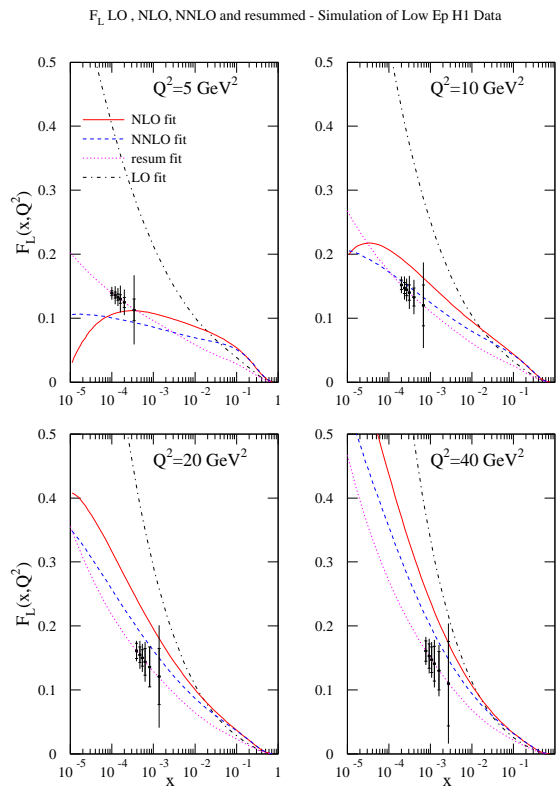}
\caption[*]{Predictions for $F_L(x,Q^2)$ at LO, NLO, NNLO and with a resummed 
calculation}
\end{center}
\end{figure}

\vspace{-1.5cm}


\end{document}